\documentclass[prb,floatfix,twocolumn,showpacs,amsmath,amssymb]{revtex4}

\usepackage{graphicx,color}
\usepackage{dcolumn}
\usepackage{bm}

\begin{document}


\title{Freezing and large time scales induced by geometrical
  frustration}

\author{Michel Ferrero,$^{1}$ Federico Becca,$^{1}$ and Fr\'ed\'eric Mila$^{2}$}

\affiliation{
  ${^1}$ INFM-Democritos, National Simulation Center, and SISSA, 
  I-34014 Trieste, Italy. \\
  ${^2}$ Institut de Physique Th\'eorique, Universit\'e de Lausanne,
  CH-1015 Lausanne, Switzerland. }

\date{\today}

\begin{abstract}
We investigate the properties of an effective Hamiltonian with
competing interactions involving spin and chirality variables,
relevant for the description of the {\it trimerized} version of the
spin-$1/2$ {\it kagome} antiferromagnet. Using classical Monte Carlo
simulations, we show that remarkable behaviors develop at very low
temperatures.  Through an {\it order by disorder} mechanism, the
low-energy states are characterized by a dynamical freezing of the
chiralities, which decouples the lattice into ``dimers'' and
``triangles'' of antiferromagnetically coupled spins.  Under the
presence of an external magnetic field, the particular topology
of the chiralities induces a very slow spin dynamics, reminiscent of
what happens in ordinary spin glasses.
\end{abstract}


\maketitle

\section{introduction}

The understanding and the characterization of the physical properties
of frustrated magnetic systems is a central issue in modern theory of
condensed matter.  Indeed, in the last years, it has become evident
that the magnetic interactions can give rise to a large variety of
different unconventional phenomena, and that they can also
be related to other phases of matter, like, for
instance, superconductivity.

From an experimental point of view, an impressive number of materials with
unconventional magnetic properties has been discovered, including
quasi-one-dimensional and ladder systems,~\cite{boucher,rice}
quasi-two-dimensional compounds with highly frustrated
magnetic interactions,~\cite{carretta}
systems of weakly coupled dimers in various 
geometries,~\cite{kageyama,kodama,rice2,rice3} and fully
three-dimensional frustrated systems based on corner sharing
tetrahedra.~\cite{pyrochlore}

\begin{figure}
 \vspace{0.3cm}
 \includegraphics[width=0.45\textwidth]{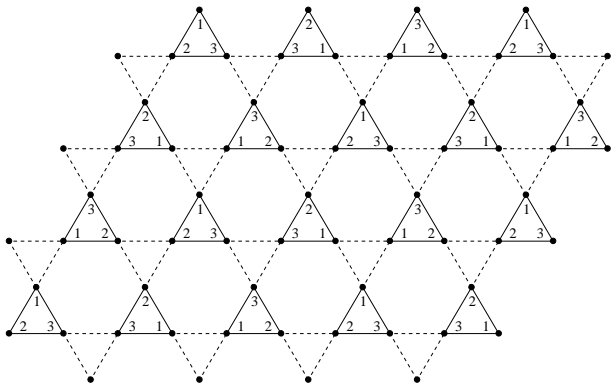}
 \caption{\label{fg:kagomedimer}
    The {\it trimerized} {\it kagome} lattice. The solid and dashed lines
    indicate the antiferromagnetic coupling $J$ and $J^\prime$,
    respectively. The numbers 1, 2 and 3 indicate the site indexing
    inside the elementary triangles which defines the gauge (see
    section~\ref{model}).}
\end{figure}

A particular role in the list of frustrated magnetic systems is played
by the magnetoplumbite ${\rm SrCr_{8-x}Ga_{4+x}O_{19}}$, in which the
anomalous magnetic properties are commonly 
attributed to planes of ${\rm Cr^{3+}}$ ions
on the so-called {\it kagome} lattice, a lattice which can be 
described as a triangular lattice of triangles 
(see Fig.~\ref{fg:kagomedimer}). In this compound,
although a small quantity of disorder is present in the Chromium planes,
there are evidences that the physical properties are not much affected
by the presence of impurities.  Nonetheless, linear-susceptibility
measurements did not find any sign of conventional magnetic order down
to $T \approx 4K$,~\cite{obradors} despite a significant
antiferromagnetic exchange coupling $J \approx 100K$ between 
${\rm Cr^{3+}}$ ions, deduced from the Curie temperature. On the other
hand, important insight into a possible spin-glass phase at very low
temperatures comes from the non-linear susceptibility $\chi_3$
measurements by Ramirez and collaborators.~\cite{ramirez} In usual
spin-glass systems, it is known that 
$\chi_3^{-1} \approx (T-T_g)^\gamma$, where $T_g$ is the spin-glass 
freezing temperature
and $\gamma$ is typically between $1$ and $4$.~\cite{tholence} In
Ref.~\onlinecite{ramirez}, it was found that $\chi_3$ diverges at 
$T_g \approx 3.3K$, with $\gamma \approx 2$, indicating a possible freezing
transition at this temperature.  It is worth noting that the
divergence of $\chi_3$ at a finite temperature indicates the
importance of three-dimensional effects, since two-dimensional
spin glasses are not expected to order at finite
temperature.~\cite{dekker}

From a theoretical point of view, the possibility of a spin-glass
behavior in the spin-$1/2$ Heisenberg model on the {\it kagome} lattice
was suggested by Chandra and collaborators.~\cite{chandra2} The
{\it kagome} lattice exhibits both frustration and a low coordinance
number, and, therefore, the Heisenberg antiferromagnet on the {\it kagome}
lattice is a good candidate to find unconventional magnetic behaviors
at very low temperature.  The most problematic issue in the {\it kagome}
antiferromagnet comes from the infinite number of classically
degenerate ground states.~\cite{chalker} As a consequence,
the linear spin-wave spectrum
of magnetic excitations possesses a whole branch of zero
modes.~\cite{harris} Moreover, exact diagonalizations on finite-size
systems have shown that the spin-spin correlations decrease very
rapidly with distance.~\cite{kag1,kag2,kag3} In addition, the series
expansion from the Ising limit and high-temperature series point to
the absence of magnetic order,~\cite{kag4,kag5} whereas, according to
large-$N$ approaches, the ground state should be disordered with unbroken
symmetry.~\cite{kag6}

The huge degeneracy of the classical ground states for the {\it kagome}
antiferromagnet opens the possibility of the so-called {\it order by disorder} 
effect.~\cite{villain} Indeed, it is known that both
thermal and quantum fluctuations can lift classical
degeneracies in favor of particular states, which break some symmetry.
Examples of this phenomenon are the ordering effect in the
antiferromagnetic Ising model on the f.c.c. lattice due to thermal
fluctuations~\cite{villain} or the establishment of a collinear order
in the $J_1{-}J_2$ model due to quantum and thermal
fluctuations.~\cite{chandra}

In particular, thermal fluctuations could be effective to partially
lift the classical ground state degeneracy by favoring states with
largest entropy. In highly frustrated antiferromagnets, like the
{\it kagome} lattice, the resulting energy landscape could still be very
complicated, with many different ``valleys'' separated by high energy
barriers, possibly giving the system a spin-glass behavior.  In
ordinary spin-glass materials, the disordered nature of the system
gives rise to frustrated magnetic interactions and, therefore, to many
topologically different configurations that are energetically
equivalent but separated by large (infinite) free energy barriers.  The
resulting physical properties show peculiar behaviors, with a very
slow spin dynamics.~\cite{spinglass,binder}

For the classical {\it kagome} model with full rotation symmetry in
spin space, no evidence in favor of a spin-glass behavior could
be found,~\cite{chalker,reimers} and it seems that one has to go 
away from this very
minimal model to have a chance to get a spin-glass-like 
behavior. Since ${\rm SrCr_{8-x}Ga_{4+x}O_{19}}$ is made
of spins 3/2, it seems natural to turn to the quantum version of the
model. In this paper, we consider an approximate way of including quantum
fluctuations into a classical description. The starting point is the
effective Hamiltonian obtained in Ref.~\onlinecite{mila} for the
spin-$1/2$ Heisenberg model on a {\it trimerized} {\it kagome} lattice
(see Fig.~\ref{fg:kagomedimer}), which contains both spin and
chirality variables.  This is a modified version of the spin-$1/2$
Heisenberg model on the {\it kagome} lattice, in which the exchange
couplings take two different values $J$ and $J^\prime$ according to
the pattern of Fig.~\ref{fg:kagomedimer}. It is worth noting that this
is actually the relevant case for the description of 
${\rm SrCr_{8-x}Ga_{4+x}O_{19}}$, since the presence of a triangular
lattice between pairs of {\it kagome} layers leads to two kinds of bond
lengths.

In Ref.~\onlinecite{mila}, by using an appropriate mean-field
approximation, this effective Hamiltonian has been useful to explain a
number of important properties of the low-lying excitations of the
{\it kagome} lattice, in particular the increase of the low-energy singlet
states with the number of sites $N$ like $1.15^N$.  
Unfortunately, this effective
Hamiltonian is far from being fully investigated because the
interactions couple the spins and the
chiralities in a non-trivial way.  Nevertheless, given the very
complicated nature of the problem, even at the classical level, the
interplay between spins and chiralities can give rise to an
unconventional behavior.  In
this respect, for any calculation in the true quantum case, it is very
important to know the underlying classical states, and, therefore, a
first fundamental step is to establish the classical picture.

Recently, an interesting step toward the comprehension of the
interplay between two different kinds of classical variables has been
achieved by using a simple Ashkin-Teller model
on a square lattice.~\cite{miladean} It was pointed out that the
presence of two local degrees of freedom might be a key element to
prevent any kind of long-range order and to obtain a very slow spin
dynamics (reminiscent of a spin-glass-like behavior) at low-enough
temperature.  Hopefully, this finding can be seen as a first and crude
explanation of the low-temperature behavior of 
${\rm SrCr_{8-x}Ga_{4+x}O_{19}}$, which does not show any magnetization
down to very low temperature, but presents some promising evidence for
a spin-glass behavior.  It is also worth noting that, although in
${\rm SrCr_{8-x}Ga_{4+x}O_{19}}$ the ${\rm Cr^{3+}}$ ions have spin
$S=3/2$, we consider the case treated in Ref.~\onlinecite{mila},
suitable for $S=1/2$.~\cite{note}

In order to go beyond the simple model considered in
Ref.~\onlinecite{miladean}, based on Ising variables, and to further
clarify the possibility to obtain a spin-glass behavior in a more
realistic system that does not contain explicit disorder but only
frustration between two distinct local variables, we investigate the
effective model of Ref.~\onlinecite{mila} in the classical limit.

The paper is organized as follows: In section~\ref{model}, we
introduce the model, in section~\ref{zerofield} we present the results
for the case of zero external magnetic field, in section~\ref{field}
we present the results of the magnetization in the presence of an
external magnetic field, and in section~\ref{conclusions} we
discuss the results and give our conclusions.

\section{the model}\label{model}

Our starting point is the effective model obtained in
Ref.~\onlinecite{mila} for the quantum spin-$1/2$ antiferromagnetic
Heisenberg model on the {\it trimerized} {\it kagome} lattice.  In this
modified version of the usual spin-$1/2$ {\it kagome} antiferromagnet, the
lattice is decomposed such as to emphasize the role of the elementary
triangles building the {\it kagome} lattice. Indeed, the spins are coupled
through two different constants $J$ and $J^\prime$ as depicted in
Fig.~\ref{fg:kagomedimer}.  In the limit $J^\prime/J \ll 1$, the
lattice can be seen as a triangular lattice of weakly coupled
triangles and, by using ordinary perturbation theory, it is possible
to derive an effective Hamiltonian in the subspace of the ground
states of the triangles. The four-fold degenerate 
ground state of a triangle can
be described by two spin-$1/2$ degrees of freedom, a spin 
${\vec \sigma}$ and a chirality ${\vec \tau}$. With these variables, the
effective Hamiltonian is defined on a triangular lattice of $N$ sites
and reads:~\cite{mila}
\begin{eqnarray}
  {\cal H}_0^{{\rm eff}} &=& J^\prime
  \sideset{}{^\prime}\sum_{\langle i,j \rangle}
  {\cal H}_{ij}^\sigma {\cal H}_{ij}^\tau, 
  \label{eq:hamilt}  \\
  {\cal H}_{ij}^\sigma &=& {\vec\sigma}_i \cdot {\vec\sigma}_j, \nonumber \\
  {\cal H}_{ij}^\tau &=& \frac{1}{9}
  [1 - 2 (\alpha_{ij} \tau_i^- + \alpha_{ij}^2 \tau_i^+)]
  [1 - 2 (\beta_{ij} \tau_j^- + \beta_{ij}^2 \tau_j^+)], \nonumber
\end{eqnarray}
where $\sideset{}{^\prime}\sum_{\langle i,j \rangle}$ denotes the sum
over pairs of nearest-neighbor
triangles. The complex parameters $\alpha_{ij}$ and $\beta_{ij}$
depend on the type of bond: They take the values $1$, $\omega$ or
$\omega^2$ [$\omega=\exp(2\pi i/3)$] when the original spins in
triangles $i$ and $j$ involved in the bond ($i,j$) sit at site $1$,
$2$ or $3$ respectively, with the convention of
Fig.~\ref{fg:kagomedimer}. Note that we have the freedom to choose
the numbering of the sites within each triangle, and that the convention
of Fig.~\ref{fg:kagomedimer} is different from that of Ref.~\onlinecite{mila}. 

With a little algebra, we write the effective Hamiltonian in a more
compact way. Indeed, the operators $\tau_i^+$ and $\tau_i^-$ can be
rewritten by introducing $\tau_i^x$ and $\tau_i^y$:
\begin{eqnarray}
  \tau_i^+ &=& \tau_i^x + i \tau_i^y, \nonumber \\
  \tau_i^- &=& \tau_i^x - i \tau_i^y. \nonumber
\end{eqnarray}
Hence, $\alpha_{ij} \tau_i^- + \alpha_{ij}^2 \tau_i^+ =
(\alpha_{ij} + \alpha_{ij}^2) \tau_i^x + i 
(\alpha_{ij}^2 - \alpha_{ij}) \tau_i^y$, and therefore:
\begin{equation}
  \alpha_{ij} \tau_i^- + \alpha_{ij}^2 \tau_i^+ =
  \begin{cases}
    2 \tau_i^x &
    \text{if $\alpha_{ij} = 1$} \\
    - \tau_i^x - \sqrt{3} \tau_i^y &
    \text{if $\alpha_{ij} = \omega^2$} \\
    - \tau_i^x + \sqrt{3} \tau_i^y & \text{if $\alpha_{ij} = \omega$.}
  \end{cases} \nonumber
\end{equation}
The notation is further simplified by introducing three unitary vectors
${\vec e}_1=(1,0)$, ${\vec e}_2=(-\frac{1}{2},-\frac{\sqrt{3}}{2})$,
${\vec e}_3=(-\frac{1}{2},\frac{\sqrt{3}}{2})$:
\begin{equation}
  \alpha_{ij} \tau_i^- + \alpha_{ij}^2 \tau_i^+ =
  \begin{cases}
    2 {\vec \tau}_i \cdot {\vec e}_1 &
    \text{if $\alpha_{ij} = 1$} \\
    2 {\vec \tau}_i \cdot {\vec e}_2 &
    \text{if $\alpha_{ij} = \omega^2$} \\
    2 {\vec \tau}_i \cdot {\vec e}_3 &
    \text{if $\alpha_{ij} = \omega$}.
  \end{cases} \nonumber
\end{equation}

\begin{figure}
 \vspace{0.3cm}
 \includegraphics[width=0.45\textwidth]{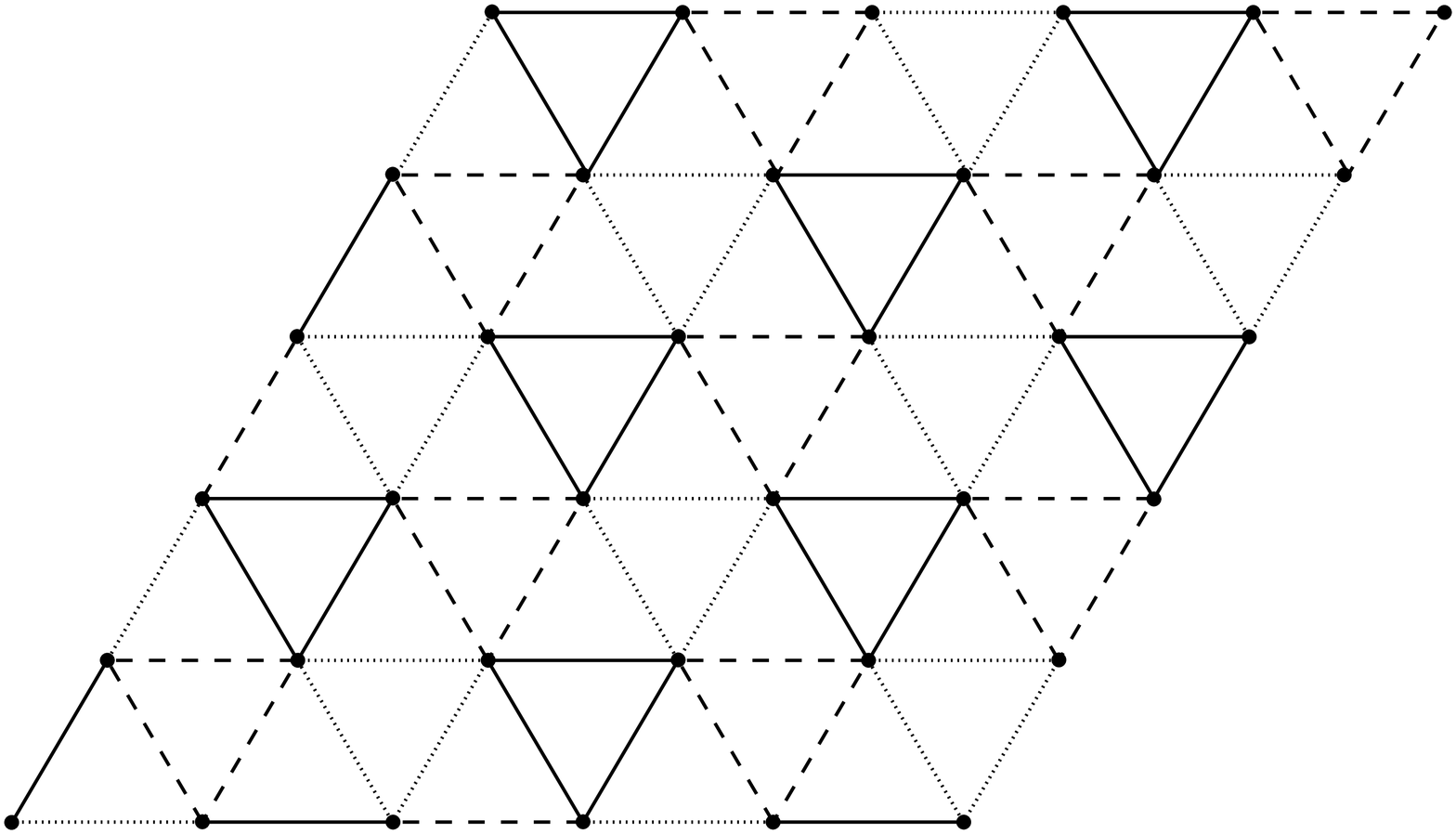}
 \caption{\label{fg:triankind}
    Triangular lattice on which the effective Hamiltonian is defined.
    The unitary vector for the bond is indicated by solid lines
    (${\vec e}_\mu = {\vec e}_1$), dashed lines (${\vec e}_\mu = {\vec
      e}_2$), and dotted lines (${\vec e}_\mu = {\vec e}_3$).}
\end{figure}

With this notation, the Hamiltonian reads:
\begin{equation}
  {\cal H}_0^{{\rm eff}} = \frac{J^\prime}{9}
  \sideset{}{^\prime}\sum_{\langle i,j \rangle}
  {\vec\sigma}_i \cdot {\vec\sigma}_j
  (1 - 4 {\vec e}_{ij} \cdot {\vec \tau}_i)
  (1 - 4 {\vec e}_{ij} \cdot {\vec \tau}_j),
\end{equation}
where the vectors ${\vec e}_{ij}$ are to be chosen among ${\vec e}_1$, 
${\vec e}_2$, and ${\vec e}_3$. In this Hamiltonian, we treat 
${\vec \sigma}_i$ and ${\vec \tau}_i$ as classical vectors with norm $1$
and $1/2$, respectively.  By adopting the gauge depicted in
Fig.~\ref{fg:kagomedimer}, we end up with a triangular lattice having
each bond characterized by a vector ${\vec e}_\mu$, with $\mu=1,2,3$.
In Fig.~\ref{fg:triankind}, the different lines indicate which 
${\vec e}_\mu$ has to be taken for ${\vec e}_{ij}$ in the effective
Hamiltonian.

Hereafter, we consider the classical limit of the effective
Hamiltonian and we prefer to normalize both spin and chirality to
unity, which leads to
\begin{equation} \label{eq:basichamilt}
  {\cal H}_0^{{\rm eff}} = \frac{J^\prime}{9}
  \sideset{}{^\prime}\sum_{\langle i,j \rangle}
  {\vec \sigma}_i \cdot {\vec \sigma}_j
  (1 - 2 {\vec e}_{ij} \cdot {\vec \tau}_i)
  (1 - 2 {\vec e}_{ij} \cdot {\vec \tau}_j).
\end{equation}
In this Hamiltonian, the presence of the chiralities has the effect of
changing the coupling between spins. We will see that this leads to a
complicated interplay between spins and chiralities and gives rise to
a very rich physics in the low-temperature regime.

\section{zero-field properties}\label{zerofield}

In this section we present the results of our model when no external
magnetic field is applied. We provide evidences that, for low-enough
temperatures, the system undergoes a partial {\it order by disorder} 
transition to a state where the chiralities are essentially frozen
in a given configuration.

\subsection{Ground state manifold}

In order to investigate the interplay between spins and chiralities,
let us first study the low-temperature properties of the system. This
is achieved by performing classical Monte Carlo simulations using a
heat-bath algorithm.~\cite{young} In the context of our model, the
heat-bath algorithm is efficient down to very low temperatures
($T/J^\prime \approx 10^{-6}$), and, therefore, it is possible to have
insight into very low-energy phases.  The details of this algorithm
are discussed in Appendix~\ref{montecarlo}.  For most
calculations, we considered an $18 \times 18$ triangular lattice with
periodic boundary conditions, and we checked the conclusions on
larger sizes.  However, as we will show in the following, for all
temperatures, the physical properties are governed by short-range
correlations and, therefore, the $18 \times 18$ lattice is enough to
obtain insight into the thermodynamic limit.

The first important outcome is that the energy per site converges to
$\epsilon=-J^\prime/2$ when $T \rightarrow 0$.  The particularly
simple value of the ground state energy deserves some further
investigation.  If we look back at the Hamiltonian defined by
Eq.~(\ref{eq:basichamilt}), we can find several classes of states that
have energy $\epsilon=-J^\prime/2$. For instance, when all the
chiralities are parallel to a given vector ${\vec \gamma}$ in the $XY$
plane defined by the vectors ${\vec e}_\mu$, then the two factors
involving the chiralities in Eq.~(\ref{eq:basichamilt}) become equal,
leading to a non-negative coupling between the spins, and the energy
is minimized by a $120^\circ$ N\'eel state for the spin vectors. Then,
the energy for an $N$-site system becomes:
\begin{equation}
  E = - \frac{J^\prime}{18} N \sum_{\mu=1,2,3} 
  (1 - 2 {\vec \gamma} \cdot {\vec e}_\mu)^2 = - \frac{J^\prime}{2} N,
\end{equation}
{\it independently} of the direction $\vec \gamma$ of the chiralities in
the $XY$ plane. Therefore, we obtain a large class of states which
have, beside the usual rotational symmetry of the N\'eel state of the
spins ${\vec \sigma}$, an extra continuous symmetry in ${\vec \tau}$ space.

Another important class of states is defined by those
configurations that have chiralities anti-parallel to the vectors
${\vec e}_\mu$, independently on each site. In this case, the coupling
of one site to four out of its six neighbors vanishes. Indeed, we have
that ${\vec e}_\mu$ is equal to ${\vec e}_1$ for the bond with two
neighbors, ${\vec e}_2$ with two other neighbors and ${\vec e}_3$ for
the last two bonds (see Fig.~\ref{fg:triankind}). For any choice of
${\vec \tau}_k$ among the three possible $-{\vec e}_\mu$, 
the effective magnetic coupling vanishes for
the four bonds where ${\vec \tau}_k \cdot {\vec e}_{kj} =
\frac{1}{2}$ since it contains the factor
$(1 - 2 {\vec e}_{kj} \cdot
{\vec \tau}_k) (1 - 2 {\vec e}_{kj} \cdot {\vec \tau}_j)$.  
Moreover, having taken ${\vec \tau}_k = -{\vec e}_\nu$,
the effective magnetic coupling with the remaining two neighbors is
non-zero only if ${\vec \tau}_j$ is equal to $-{\vec e}_\nu = {\vec \tau}_k$. 
In this case, the effective interaction between the two
sites is equal to $J^\prime$.

\begin{figure}
 \vspace{0.3cm}
 \includegraphics[width=0.45\textwidth]{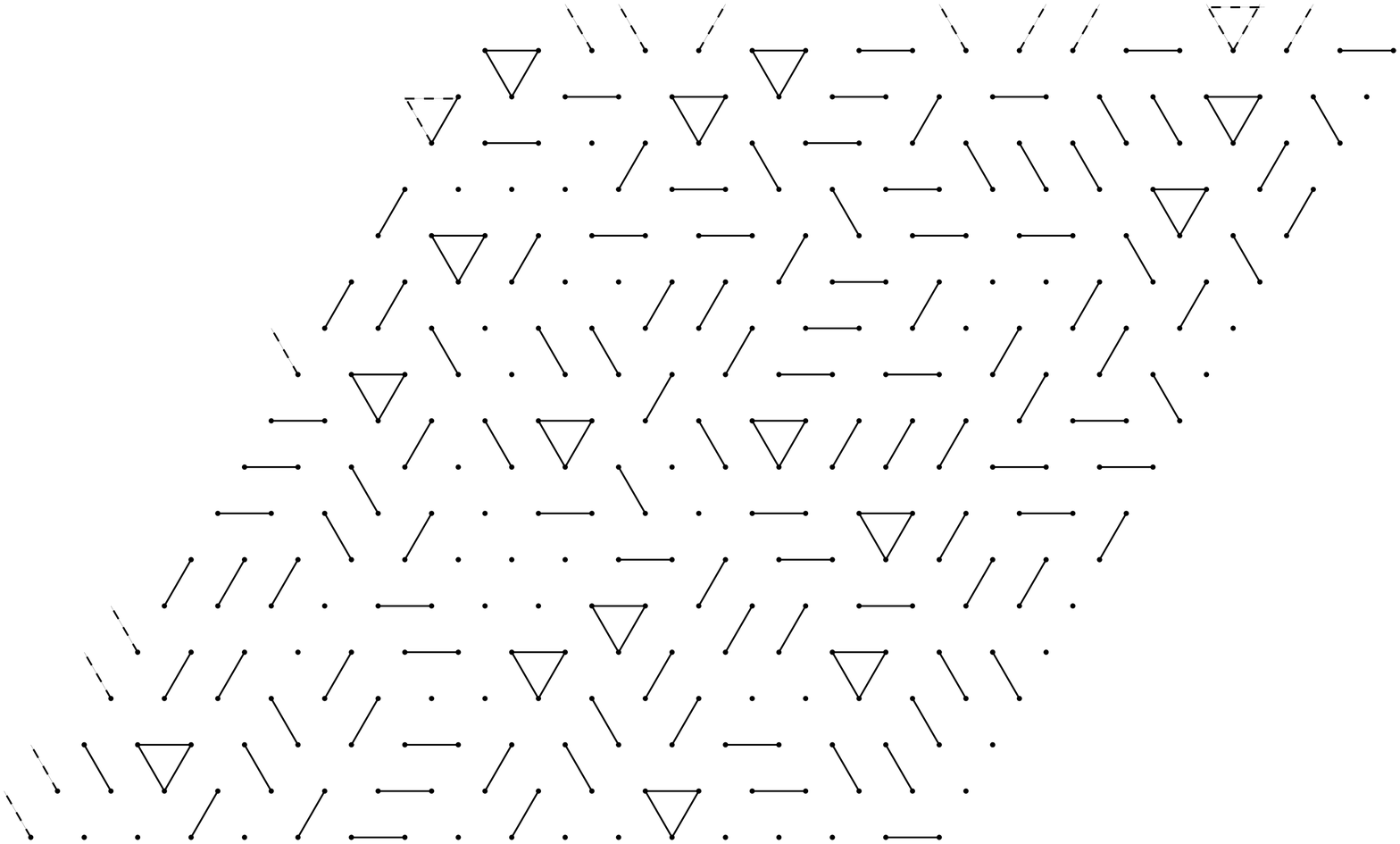}
 \caption{\label{fg:typicalcouplconf}
    Generic coupling pattern for random chiralities opposite to the
    vectors ${\vec e}_\mu$. Notice that, because of the periodic
    boundary conditions, some of the bonds (indicated by dashed lines)
    connect sites on opposite sides of the lattice.}
\end{figure}

A given site $k$ can therefore interact with its neighbors in three
different ways. It might be coupled to two neighbors, $l$ and $m$, in
which case these two neighbors will also be coupled, because they have
parallel chiralities and ${\vec e}_{lm} = {\vec e}_{kl} = {\vec e}_{km}$. 
In that way, we create a coupled triangle, which is
completely disconnected from the rest of the lattice. The site $k$
might also be coupled to one neighbor only, forming a system of two
coupled spins. The third possibility is to have the site completely
disconnected from its neighbors. A typical coupling pattern 
is shown in Fig.~\ref{fg:typicalcouplconf}. Hereafter, we
denote by dimer the system of two coupled spins. It is worth
mentioning that because of the classical nature of the spins, the
dimer is a classical object, with gapless excitations, and it has
nothing to do with the quantum dimers, which are gapped objects.

When the chiralities are opposite to the ${\vec e}_\mu$, the coupling
pattern will only exhibit isolated sites, dimers and triangles.
Moreover, any of these patterns that one can draw on the triangular
lattice can also be achieved by choosing the appropriate chiralities.
Having fixed the chiralities, the energy is minimized with opposite
spins on the dimers and the energy per spin is $\epsilon=-J^\prime/2$;
on an isolated triangle, the sum of the spins needs to be zero, the
energy per spin being again $\epsilon=-J^\prime/2$; finally isolated
sites do not contribute to the energy. Hence, among the possible
configurations of chiralities opposite to ${\vec e}_\mu$, those that
do not have isolated spins have the ground state energy per site
$\epsilon=-J^\prime/2$.

A direct inspection of the chiralities and spins in our Monte Carlo
simulations shows that when the system is cooled down abruptly from
infinite temperature, that is by starting at the chosen temperature
from a random configuration for the spins and the chiralities,
states with no obvious pattern in the chiralities and energy
per site $\epsilon=-J^\prime/2$ are achieved.  Therefore, the two
particular classes of states described above are only two among many
other ground states.  On the contrary, by cooling down the system
through a finite number of temperatures and letting it thermalize at
every step before cooling down to the next nearby temperature, we
obtain a configuration with disconnected triangles and dimers.  Within
the heat-bath Monte Carlo, this fact strongly indicates the presence of
a partial {\it order by disorder} effect: Among all the possible ground
states, the states with large entropy are selected by thermal
fluctuations. In the following, we will describe in more details
the entropic selection that appears at low-enough temperature,
describing how this transition affects the spin dynamics.

\subsection{Entropic selection of a family of ground states}

At temperature $T/J^\prime=10^{-6}$, the Monte Carlo results show that
the chiralities are frozen on each site in the $XY$ plane opposite to the
${\vec e}_\mu$ vectors in such a way as to create a collection of
disconnected dimers and triangles. The selection of this specific
class of states can be explained by its large entropy. Within each
decoupled triangle and dimer, the spins are totally free to rotate
as long as their sum vanishes {\it independently} of the spins of
the other triangles and dimers.
Thus, all those states have a huge entropy at low-enough
temperature. It can easily be proved that both for dimers and
triangles, every site carries one degree of freedom that does not
change the energy. This point is discussed in more details below
in the section devoted to the specific heat. As the system is cooled 
down slowly, these
states with largest entropy are selected. The ground state degeneracy
is partially lifted by thermal fluctuations and an 
entropic {\it order by disorder} effect is observed.

\begin{figure}
 \vspace{0.3cm}
 \includegraphics[width=0.45\textwidth]{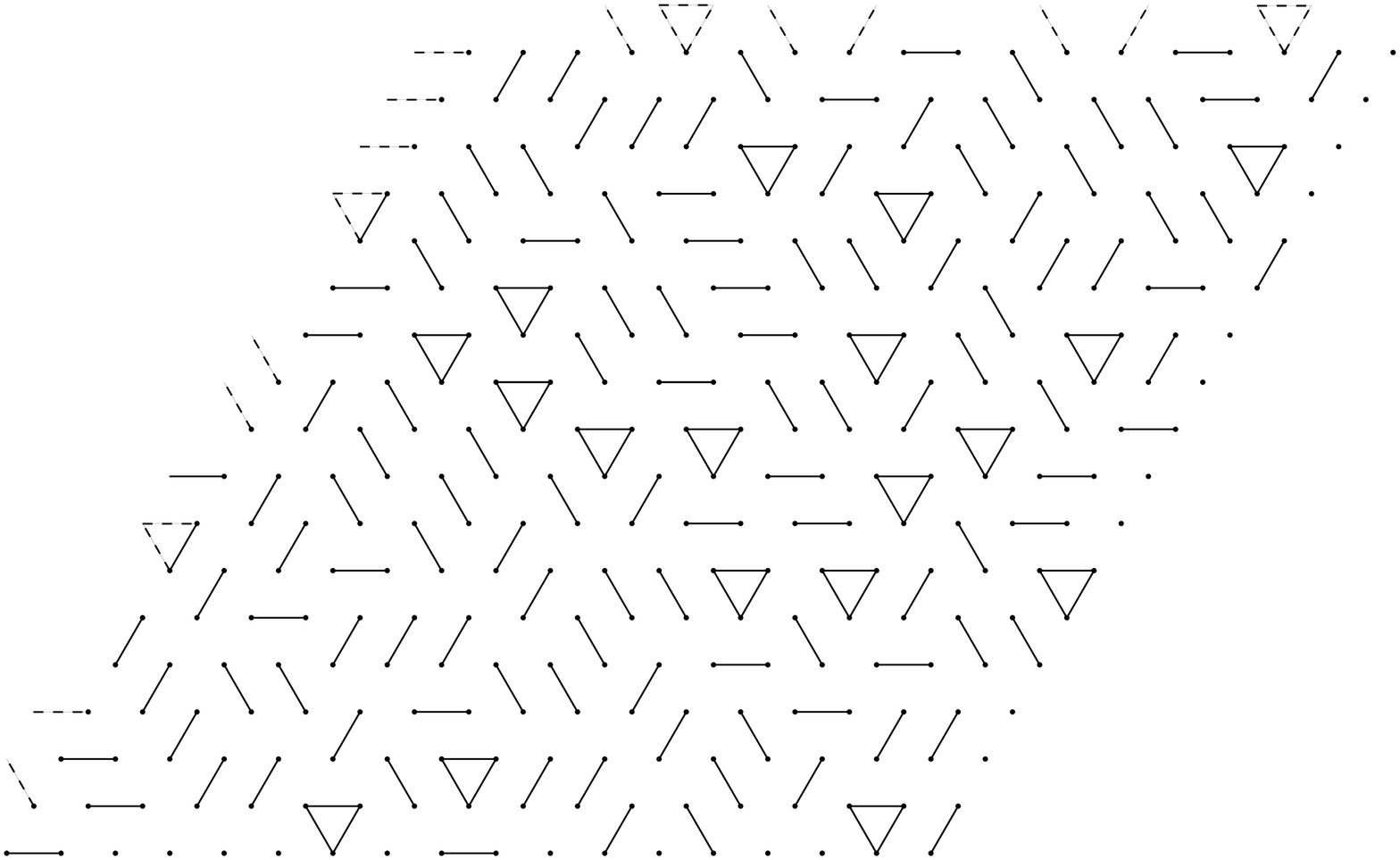}
 \caption{\label{fg:couplconf}
    Typical coupling pattern at $T/J^\prime = 10^{-6}$: The solid
    lines between the sites indicate a non-zero coupling between the
    spins.  Notice that, because of the periodic boundary conditions,
    some of the bonds (indicated by dashed lines) connect sites on
    opposite sides of the lattice.}
\end{figure}

In Fig.~\ref{fg:couplconf}, we show how the non-zero couplings are
typically distributed over the lattice at $T / J^\prime = 10^{-6}$. We
do not have any evidence that the system prefers to create triangles
or dimers: We have performed several simulations starting from different 
random orientations, and in all cases we have reached a state which
contains both triangles and dimers. The average number of triangles
is 17 on an $18 \times 18$ site cluster, and the distribution is peaked around
this number with a mean square displacement of 3. We suspect that
this simply corresponds to the distribution for all coverings of
the triangular lattice by triangles and dimers. In any case, this observation
agrees with the fact that, at low temperature, triangles and dimers have 
the same entropy per site.

\begin{figure}
 \vspace{0.3cm}
 \includegraphics[width=0.3\textwidth]{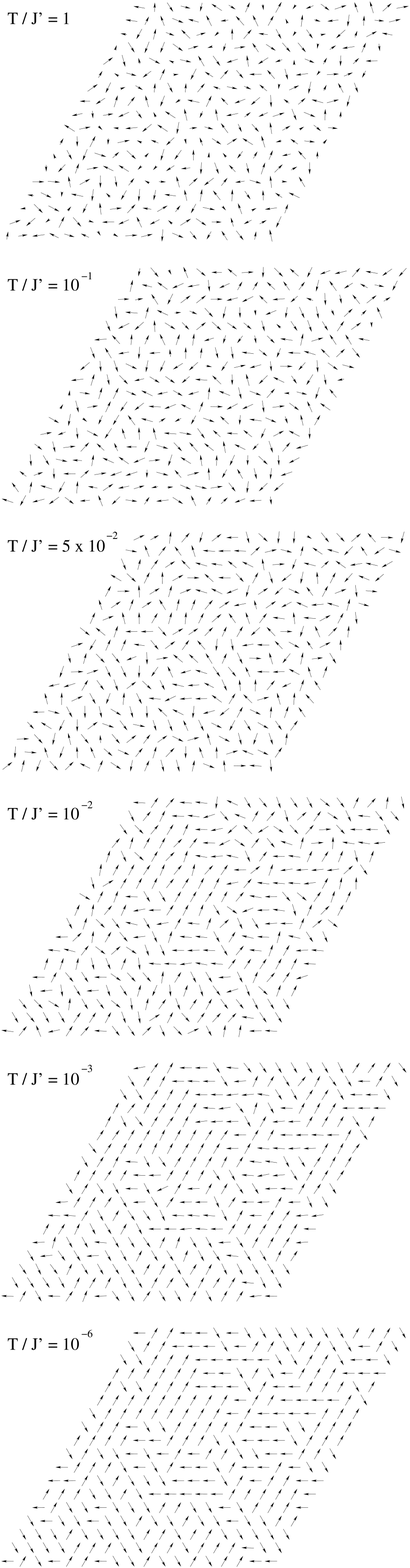}
 \caption{\label{fg:chiralconfcool}
    Configuration of the chiralities in the $XY$ plane for different
    temperatures.}
\end{figure}

To have more insight into this entropic selection, we report the
chiralities in the $XY$ plane for decreasing temperatures in
Fig.~\ref{fg:chiralconfcool}. As the system is cooled down below $T^*
\approx 0.05 J^\prime$, the chiralities start to display a short-range
order in the $XY$ plane, getting more and more anti-parallel to the
$\vec{e}_\mu$ vectors. 
As far as we can tell, this cross-over temperature does not depend
on the cooling rate, provided it is not too fast. The results reported
in Fig.~\ref{fg:chiralconfcool} have been obtained after performing $10^4$
sweeps over the lattice at each intermediate temperature starting from
a random configuration, but the same behaviors have been observed
with different numbers of sweeps.

Since the chirality configurations that can be selected by the 
{\it order by disorder}
mechanism  form an infinite but {\it  discrete} set, one can expect 
that, once such a state has been reached,  the system will 
somehow be trapped in this state. This is indeed the case on short 
time scales. However, whether this corresponds to a true freezing
transition requires a careful study of the low temperature dynamics.
For instance, in 2D models of spin glasses with short range interactions,
the actual freezing on arbitrarily large time scales only takes
place at zero temperature. So we now turn to a careful analysis
of the low-temperature spin and chirality dynamics.

\subsection{Low-temperature spin and chirality dynamics}

\begin{figure}
 \vspace{0.3cm}
 \includegraphics[width=0.45\textwidth]{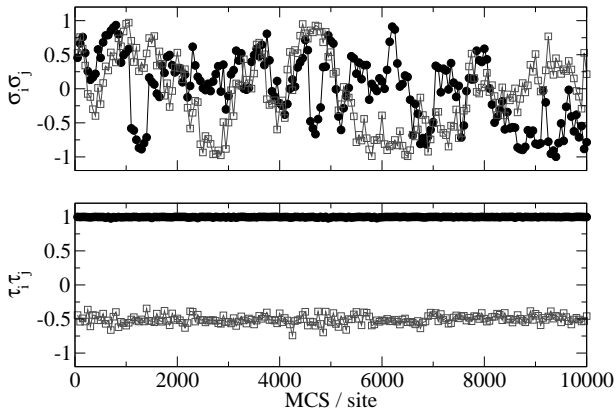}
 \caption{\label{fg:corr}
    Spin-spin (upper panel) and chirality-chirality (lower panel)
    correlations at $T/J^\prime =
    10^{-3}$ for two non-nearest-neighbor sites with parallel 
    (full symbols) and non-parallel (open symbols) chiralities.}
\end{figure}

Below $T^*$, spins and chiralities behave
very differently. Whereas the chiralities are frozen on short time
scales,
spins are expected to continue to have a fast dynamics since, 
once the lattice is
decoupled into disconnected dimers and triangles, the spins belonging
to each of these elementary building blocks are free to rotate as long
as the total spin of the dimer or of the triangle remains zero.
Spin-spin and chirality-chirality correlation functions confirm this
fact.  In Fig.~\ref{fg:corr}, we show typical non-nearest-neighbor
correlations ${\vec \sigma}_i(t) \cdot {\vec \sigma}_j(t)$ and 
${\vec \tau}_i(t) \cdot {\vec \tau}_j(t)$ as a function of the time 
after the system has reached equilibrium at $T /
J^\prime = 10^{-3}$. Throughout, the time unit is one Monte Carlo
sweep over the entire lattice. 
Indeed, the spin-spin correlation function for non-nearest-neighbors
fluctuates wildly on short time scales, while the chirality-chirality
correlation function is essentially constant.
By contrast, the spin-spin 
correlations for a
pair of coupled nearest-neighbor sites are fixed to $-1$ or $-1/2$ for
dimers or triangles, respectively (not shown).

On short time scales,
the chiralities are frozen below $T^*$ and although there is a huge
class of energetically and entropically equivalent states, once one 
configuration is
achieved, the system cannot switch to another one.  The different
configurations must therefore be separated by large energy barriers and the
chiralities remain trapped in a small region of their phase space. A similar
behavior is observed in ordinary spin-glass materials, where the
spin degrees of freedom are completely frozen below a given
temperature and different states are separated by infinite (in the
thermodynamic limit) energy barriers. In this case, the transition
temperature can be characterized by the divergence of the spin-glass
susceptibility
\begin{equation}
  \chi_{\rm SG}^\sigma = \frac{1}{N}
  \sum_{i,j} \left [ \frac{1}{t_{\rm exp}} \sum_{t=1}^{t_{\rm exp}}
  {\vec \sigma_i}(t) \cdot {\vec \sigma_j}(t) \right ]^2,
\end{equation}
where the average over the time (Monte Carlo steps) is done over a
sufficiently large period $t_{\rm exp}$, corresponding to the time scale of
the experiment. In ordinary spin glasses, 
$\chi_{\rm SG}^\sigma$ is zero (very small on finite systems) for temperatures
above the critical temperature, where the spins have a fast dynamics,
and becomes finite below the critical temperature, indicating that the
spins are frozen in a given configuration. In particular, at low
temperature, $\chi_{\rm SG}^\sigma$ grows linearly with the number of sites
$N$.

\begin{figure}
 \vspace{0.3cm}
 \includegraphics[width=0.45\textwidth]{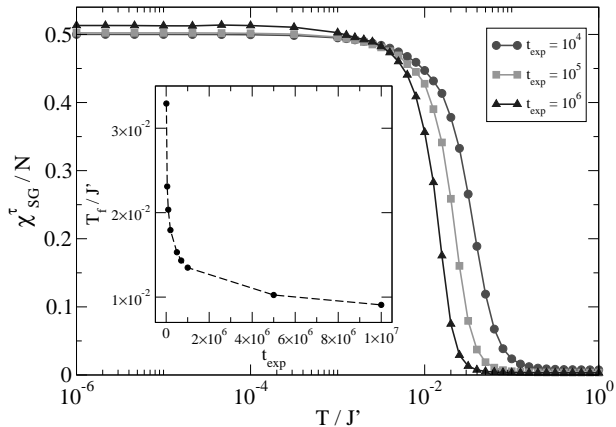}
 \caption{\label{fg:chisgtau}
    Chiral spin-glass susceptibility per site $\chi_{\rm SG}^\tau / N$
    as a function of the temperature for different time scales $t_{\rm exp}$.
    Inset: Variation of the freezing temperature $T_f$ with $t_{\rm exp}$.
    Lines are guides to the eye.}
\end{figure}

By analogy, in our model, we can define a spin-glass susceptibility
for the chirality pseudo-spins:
\begin{equation}
  \chi_{\rm SG}^\tau = \frac{1}{N} 
  \sum_{i,j} \left [ \frac{1}{t_{\rm exp}} \sum_{t=1}^{t_{\rm exp}}
  {\vec \tau_i}(t) \cdot {\vec \tau_j}(t) \right ]^2,
\end{equation}
and try to identify a transition temperature from the high-temperature 
disordered phase to the low-temperature frozen phase. It turns out that,
like in 2D disordered spin models, the freezing takes place at finite
temperatures only for finite time scales, and that the actual phase
transition, if any, presumably takes place at zero temperature.
To see this, we have plotted in Fig.~\ref{fg:chisgtau} the chirality
susceptibility $\chi_{\rm SG}^\tau$ as a function of temperature
for different {\it experimental} times $t_{\rm exp}$, corresponding 
in this case, to the time we have waited at each temperature during the
cooling process. For a given time scale, one can
clearly identify two regimes:
At high temperatures $\chi_{\rm SG}^\tau /N$ is very small,
due to the dynamical character of the chirality variables, whereas for
$T/J^\prime$ small enough, there is a clear enhancement of 
$\chi_{\rm SG}^\tau / N$, indicating a frozen dynamics of the chiralities. 
However, the temperature at which this change of behavior takes place
depends on the time of the simulation. To quantify this dependence,
we have plotted the freezing temperature $T_f$ at which the susceptibility
reaches half the value it will take at zero temperature as a function
of $t_{\rm exp}$. Clearly $T_f$ decreases with $t_{\rm exp}$, and
for times $t_{\rm exp}>10^5$, the two are related by 
$t_{\rm exp}=t_0 \exp (E_0/T_f)$, or equivalently 
$T_f=E_0/\log(t_{\rm exp}/t_0)$, with $E_0\simeq 0.07 J^\prime$ 
and $t_0\simeq 6500$.

This fit shows that, at any temperature, the chirality configuration
will change on large enough time scales, and that the process
is thermally activated with an activation energy $E_0$. This energy
can thus be interpreted as the typical free-energy barrier between
two configurations selected by the {\it order by disorder} mechanism.
As in the model of Ref.\onlinecite{miladean}, the freezing temperature
$T_f$ has only a weak logarithmic dependence on the time scale.

Note that these results are not affected by the size ($18 \times 18$) 
used in the simulation. Indeed we have checked in a number of cases
that the conclusions
are qualitatively and quantitatively similar for $27 \times 27$
and $36 \times 36$ clusters. In particular, for a given $t_{\rm exp}$,
the temperature $T_f$ does not change upon increasing the size of the
cluster.

To summarize, we found that spins and chiralities have very different
low temperature dynamics in zero field. At a cross-over temperature
$T^*\simeq 0.05 J'$, chirality configurations with maximum entropy 
are selected.
Below this temperature, the chirality dynamics is activated with a 
an energy barrier $E_0\simeq 0.07 J'$ and it is already extremely
slow at $T=0.01 J'$,
while the spins retain a very fast dynamics down to the lowest
temperatures.  
The spins do not behave as free spins however, as we shall
see in the following section, and the freezing of the chiralities has
important consequences on the response of the spins in the presence of
an external magnetic field.

\subsection{Specific heat}

\begin{figure}
 \vspace{0.3cm}
 \includegraphics[width=0.45\textwidth]{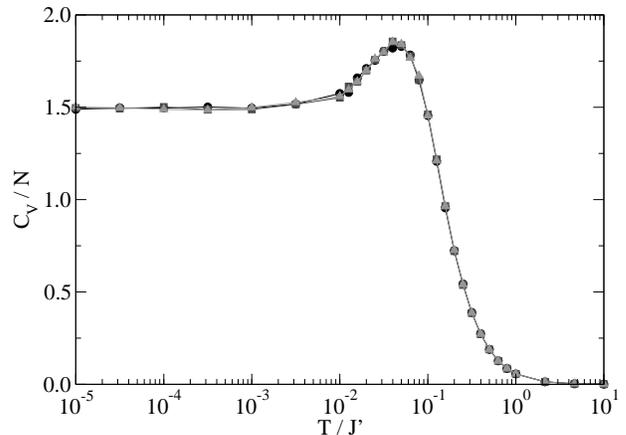}
 \caption{\label{fg:specheatgeneric}
    Specific heat $C_V$ per spin as a function of the temperature for
    different lattices: $18 \times 18$ (circles), $27 \times 27$
    (squares), and $36 \times 36$ (triangles). Lines are guides to the
    eye.}
\end{figure}

On very general grounds, we can expect that the cross-over from the
high-temperature disordered phase to the low-temperature phase
where chirality configurations have been selected by the {\it order 
by disorder} mechanism is marked by a peak in the specific heat. Whereas
for a true phase transition the specific heat diverges at the critical
temperature, for freezing phase transitions, like in spin-glass
systems, the specific heat exhibits a rather broad peak near the
freezing temperature.~\cite{binder} This is indeed what we observe in
our model. We calculate the specific heat using $C_V = (\langle E^2
\rangle - \langle E \rangle^2) / T^2$, where $E=E(T)$ is the internal
energy. The specific heat per spin shows a broad peak which does not
increase with the size of the lattice for $T\simeq T^* \simeq 0,05 J'$ 
(see Fig.~\ref{fg:specheatgeneric}). 

The low-temperature value of the specific heat, namely 3/2 per
site, is a very nice confirmation of the picture. For classical
variables, each quadratic mode gives a contribution to the 
specific heat equal to 1/2 in units of $k_B$. Thus, if both spin and chirality
were ordered at low temperature, we would expect the zero temperature
specific heat per site to be equal to 2. According to our 
picture however, only the chiralities are ordered. In a given
configuration that minimizes the entropy, each chirality can
fluctuate around its reference direction $-\vec e_\mu$ with two 
quadratic modes
corresponding to two orthogonal directions, and its contribution
to the specific heat is equal to 1. However, within a triangle
or a dimer, the spins are free to rotate as long as their sum 
is equal to zero. This means that they have zero-energy modes
which do not contribute to the specific heat. For a dimer, the
common direction of the vectors can rotate freely, while the relative
direction gives rise to two quadratic modes, so one effectively gets
one quadratic mode per spin. In a triangle, if we consider a configuration
where the sum of the spins is equal to zero, the orientation of a
given spin is arbitrary, which gives rise to two zero-energy modes,
and the others are free to rotate around the direction of the first
one, which gives a third zero-energy mode. So we are left with 3
quadratic modes, i.e. one quadratic mode per spin, like in a dimer.
So the spin contribution to the specific heat is 1/2 per site,
and the total specific heat per site is equal to 3/2. Note that the
entropy is thus the same for triangles and dimers, as stated above.

The reduction of the spin contribution to the specific heat by a factor
2 is quite remarkable. This should be compared for instance with
the classical Heisenberg model on the {\it kagome} lattice, in which 
case the zero temperature specific heat is reduced from 1 to 11/12 
(see Ref.\onlinecite{chalker}).

\section{finite-field properties}\label{field}

In the absence of an external magnetic field the chiralities show a
behavior reminiscent of spin glasses. In ordinary spin-glass materials, the
freezing of the degrees of freedom, together with a complicated energy
landscape, has consequences on the magnetic properties of the system.
Even though in our model the spins retain a fast dynamics even below
$T^*$, the freezing of the chiralities might still induce non-trivial
properties when an external magnetic field is present.  In the
following, we consider an external magnetic field that is coupled to
the spins and we study the dynamical response of the spins.

\subsection{Equilibrium magnetization}

The Hamiltonian in the presence of an external magnetic field $h$
along the $z$ axis reads:
\begin{equation}
  {\cal H}^{{\rm eff}} = {\cal H}_0^{{\rm eff}} + 
  h \sideset{}{^\prime}\sum_i \sigma_i^z.
\end{equation}
First, we concentrate on the equilibrium magnetization of the system
as a function of the temperature.  In order to find the equilibrium
magnetization per site $m$, we perform a very long simulation, cooling
down the system through a finite number of intermediate temperatures
in the presence of the external magnetic field. At each temperature,
we equilibrate the system, by considering a large number of Monte
Carlo steps: This kind of protocol is known as a {\it field-cooled}
(FC) experiment.  Within this procedure, we are able to obtain
accurate results for the equilibrium curve for the magnetization as a
function of the temperature.

\begin{figure}
 \vspace{0.3cm}
 \includegraphics[width=0.45\textwidth]{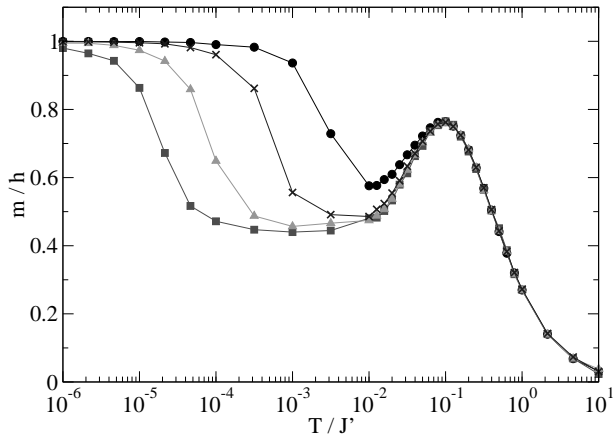}
 \caption{\label{fg:magnetvstemp}
    The equilibrium magnetization per spin in units of $h$ as a
    function of the temperature, for $h/J^\prime=10^{-2}$ (squares),
    $2 \times 10^{-2}$ (triangles), $5 \times 10^{-2}$ (crosses),
    $10^{-1}$ (circles). Lines are guides to the eye.}
\end{figure}

In Fig.~\ref{fg:magnetvstemp}, we report the results for the
magnetization per site as a function of the temperature, for different
values of the external magnetic field $h$.  We can distinguish three
different regimes. In the first one, common to all the cases and
corresponding to $T/J^\prime \gtrsim 0.1$, the chiralities are
randomly distributed and are free to rotate, retaining a quite fast
dynamics. Therefore, 
$\langle {\vec e}_{ij} \cdot {\vec \tau}_j \rangle \approx 0$, 
implying that the effective Hamiltonian reduces to
a Heisenberg model on a triangular lattice, with an effective
antiferromagnetic coupling $J^{\rm eff} \approx J^\prime/9$ [see
Eq.~(\ref{eq:basichamilt})].

In the second regime, whose temperature range strongly depends on the
value of the magnetic field, the magnetization decreases and shows a
plateau whose width depends on the actual value of the external
magnetic field.  This behavior can be explained by the establishment
of a short-range order of the chiralities, which become opposite to the
vectors ${\vec e}_\mu$, like in the $h=0$ case. In this regime, the
free energy is minimized because of the high entropy of the spins, the
physical properties are mostly determined by disconnected dimers and
triangles and the actual value of the magnetization depends on the
number of dimers and triangles, being $m=h/(2J^\prime)$ for the former
case and $m=h/(3J^\prime)$ for the latter case.

\begin{figure}
 \vspace{0.3cm}
 \includegraphics[width=0.45\textwidth]{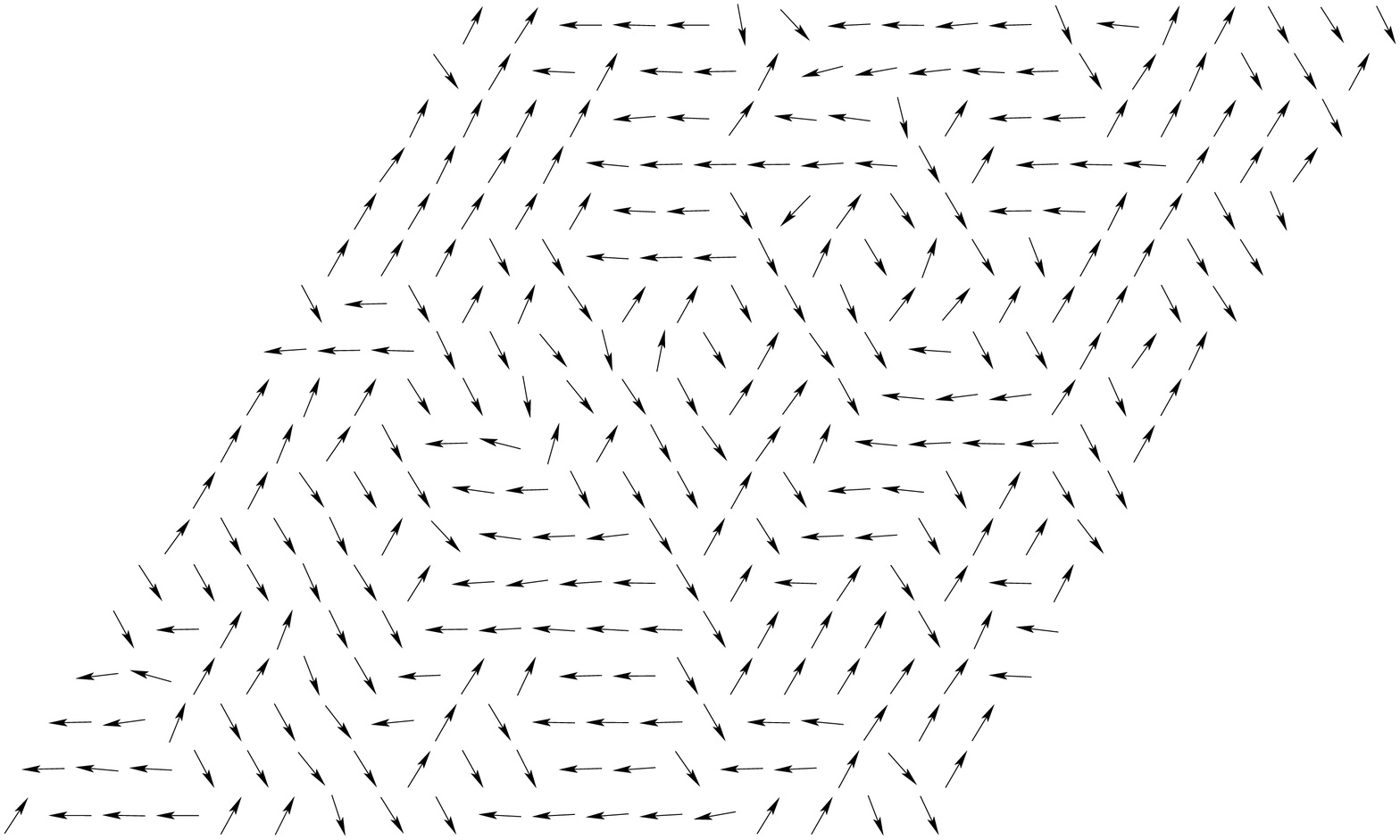}
 \caption{\label{fg:chiralmagnet}
    Configuration of the chiralities in the $XY$ plane for $T/J^\prime
    = 10^{-6}$ and an external magnetic field $h/J^\prime = 10^{-2}$.}
\end{figure}

By further lowering the temperature, the entropic part of the free
energy becomes less important and the chiralities rearrange by
slightly modifying their ordering, allowing the spins to increase
their magnetization as the system reaches its actual ground state for
finite $h$.  As $T \rightarrow 0$, the magnetization per site goes to
$m_{\rm max} = h/J^\prime$. A direct inspection of the chiralities,
see Fig.~\ref{fg:chiralmagnet}, shows that their low-temperature
configuration is only slightly modified from the $h=0$ case.  However,
this is sufficient to change considerably the magnetization.
The low temperature $T_m(h)$ at which the magnetization finally
increases indicates that there is a very small energy difference
between the actual ground state and the states described by dimers and
triangles.

\begin{figure}
 \vspace{0.3cm}
 \includegraphics[width=0.45\textwidth]{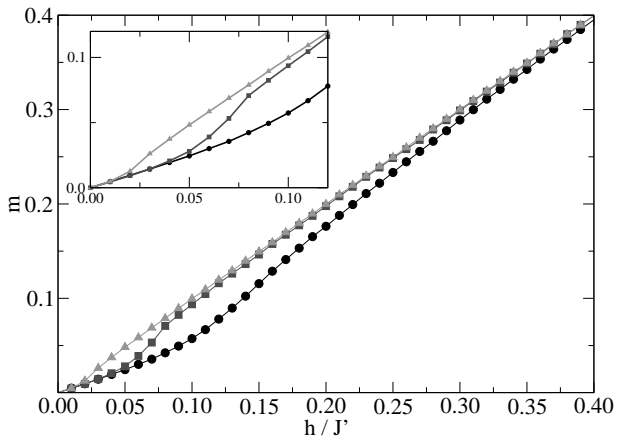}
 \caption{\label{fg:magnetvsh}
    Magnetization per site as a function of the external field at
    $T/J^\prime=10^{-4}$ (triangles), $T/J^\prime=10^{-3}$ (squares)
    and $T/J^\prime=10^{-2}$ (circles).  The lines are guides to the eye.}
\end{figure}

As shown in Fig.~\ref{fg:magnetvstemp}, the qualitative behavior
of the equilibrium magnetization is the same for different values of
$h$.  However, as the external field gets bigger, the magnetization
starts increasing towards its saturation value for higher temperatures.  
Therefore,
$T_m(h)$ strongly depends upon the external magnetic field and
increases with $h$.  On the contrary, the formation of dimers and
triangles, determined by the maximum of the magnetization at
$T/J^\prime \sim 0.1$, does not depend on the external magnetic
field, indicating that $T^*$ does not change with the application of
$h$.  The important fact is that, at a given temperature below $T^*$,
the magnetization is no longer a linear function of the applied
external field, as typical magnetization curves show in
Fig.~\ref{fg:magnetvsh}.

For small magnetic fields, the system is in the regime with
disconnected dimers and triangle, which have a linear magnetization
$m=h/(2J^\prime)$ and $m=h/(3J^\prime)$, respectively. Therefore, the
total magnetization is also linear in $h$ with a coefficient that
depends on the actual number of triangles and dimers. For larger $h$,
the magnetization goes through a highly non-linear region,
characterized by a complicated interplay between spins and
chiralities. In this regime, the entropy is maximized with chiralities
opposite to the vectors ${\vec e}_\mu$, whereas the energy is
minimized with configurations that lead to a bigger magnetization.
Eventually, the chiralities order in such a way as to allow the spins 
to have a
magnetization $m=h/J^\prime$ and the system is again in a linear
regime. A consequence of this competition between different states is
visible in the dynamical properties of the system.

\subsection{Dynamical magnetization and slow spin dynamics}

\begin{figure}
 \vspace{0.3cm}
 \includegraphics[width=0.45\textwidth]{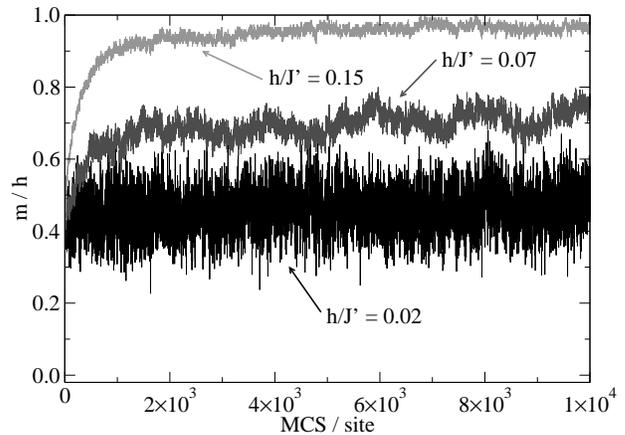}
 \caption{\label{fg:magnetvstime}
    Magnetization per site in units of the external field as a
    function of the Monte Carlo steps for $h/J^\prime=0.02$,
    $h/J^\prime=0.07$ and $h/J^\prime=0.15$.  The temperature is
    $T/J^\prime=10^{-3}$.}
\end{figure}

In order to investigate the interplay between spins and chiralities
and its effect on the dynamical properties of the system, we perform a
{\it zero-field cooled} (ZFC) numerical experiment: The system is slowly 
cooled
down to the temperature of interest in the absence of an external
magnetic field.  Then, an external magnetic field is turned on and we
observe the magnetization as a function of time.  
Fig.~\ref{fg:magnetvstime} shows the result for three
different external fields $h/J^\prime=0.02$, $h/J^\prime=0.07$ and
$h/J^\prime=0.15$ at $T/J^\prime=10^{-3}$. At equilibrium, the first
field will bring the system into a regime with disconnected dimers and
triangles and magnetization $h/(3J^\prime) \le m \le h/(2J^\prime)$
(corresponding to the plateau region of Fig.~\ref{fg:magnetvstemp}).
The third field will generate a state with chiralities slightly
different from the opposite vectors ${\vec e}_\mu$ and magnetization
$m \sim h/J^\prime$ (corresponding to the bump of
Fig.~\ref{fg:magnetvstemp} at very low temperatures), whereas the
second field represents the intermediate regime.

For $h/J^\prime=0.02$ we observe that the system acquires its final
magnetization in a very small number of Monte Carlo steps, through a very
fast process, typical of non-frustrated systems. This can be
understood because, just before turning on the magnetic field, the
system already displays a collection of dimers and triangles.  When
the external field is turned on, the spins quickly arrange to minimize
the energy. In this process, the chiralities do not change from their
previous equilibrium position.  On the contrary, when a field
$h/J^\prime=0.07$ or $h/J^\prime=0.15$ is switched on, the final
equilibrium magnetization is reached only after a much longer
simulation time.  As in the previous case, the system very quickly
acquires a magnetization corresponding to the magnetization of a
system of triangles and dimers. However, after this first very rapid
process, a much slower dynamics takes place, during which the
chiralities and spins find a compromise to minimize the free energy.
Indeed, as soon as the chiralities start to couple the triangles and
dimers with their neighbors, the spins are no longer free to move
independently and they start to have a slower dynamics. In this case
the magnetization obtained after a short time is different from the
equilibrium magnetization.

\begin{figure}
 \vspace{0.3cm}
 \includegraphics[width=0.45\textwidth]{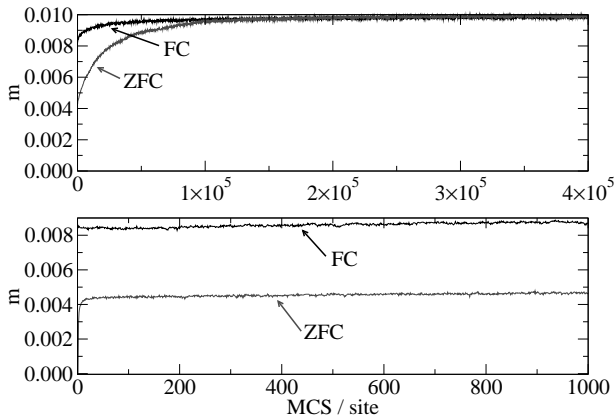}
 \caption{\label{fg:magnetvstime2}
    Long-time (upper panel) and short-time (lower panel) behavior of
    the magnetization per spin versus time at $T/J^\prime = 10^{-6}$
    for a FC and a ZFC protocol. The external field has magnitude
    $h/J^\prime = 10^{-2}$.}
\end{figure}

A further confirmation of the very slow spin response to a magnetic
field is given by the difference between the magnetization curve as a
function of the time for the FC and the ZFC protocols, see
Fig.~\ref{fg:magnetvstime2}. In the following, we concentrate on the
case $T/J^\prime=10^{-6}$ and $h/J^\prime=10^{-2}$.  Let us first
consider the ZFC magnetization. As long as $h=0$, the magnetization
vanishes and the chiralities are all opposite to the vectors ${\vec e}_\mu$ 
and there are both triangles and dimers in the coupling
pattern. When the external magnetic field is turned on, the
magnetization reaches the value $m \approx 0.004$ in about $20$ Monte
Carlo steps.  This first very rapid regime corresponds to the spins
moving toward the direction of the magnetic field. During this fast
spin dynamics, the chirality does not play any important role in the
dynamical process.  Indeed, we have verified that $m \sim 0.004$ is
also the magnetization acquired in the first $20$ Monte Carlo steps
within a simulations where the chiralities were kept fixed.  In the
second regime, the magnetization increases very slowly up to its
saturation value $m \sim h/J^\prime$. This slow increase is due to the
combined dynamics of spins and chiralities, in order to find a
compromise to minimize the free energy.
By contrast, within the FC protocol, the system is cooled down in the
presence of the external magnetic field and therefore, at the final
temperature, the magnetization has already a finite value $m \sim
0.0085$ and the chiralities are no longer in the direction of the
basis vectors. Then, by keeping the system at the desired temperature
$T < T_m(h)$, a very slow increase of the magnetization is observed,
similar to the behavior of the second regime of ZFC.

The pronounced difference between the ZFC and the FC protocols, and
more generally an evident {\it aging} effect has already been
emphasized in a related, but much simpler classical model containing
two different Ising variables.~\cite{miladean} Finding the same kind
of behavior in a more realistic Hamiltonian demonstrates that this is
a robust feature of models with competing variables and frustrating
interactions, even in the absence of an explicit disorder like in
ordinary spin-glass materials.

\section{conclusions}\label{conclusions}

In this paper we have investigated the properties of the classical version of
an effective Hamiltonian for the $S=1/2$ Heisenberg antiferromagnet on a
{\it trimerized} {\it kagome} lattice using classical Monte Carlo methods.
The analysis of this model shows many interesting features, possibly
reminiscent of a spin-glass behavior at low temperatures.  The fully
frustrated character of the interaction, due to the presence of two
kinds of variables on each site, induces a freezing of the chiralities
and, in the presence of an external magnetic field, the development 
of very large time scales for the spins at low temperature.
Indeed, we have demonstrated that,
by a partial {\it order by disorder} effect, for $T \lesssim T^* \approx
0.05J^\prime$, the states with highest entropy are selected by thermal
fluctuations, and the chiralities freeze along particular directions.
This prevents the system from developing any long-range correlations and
completely splits the lattice into disconnected dimers and triangles,
with antiferromagnetically coupled spins. On each of these elementary
entities the spins are free to rotate, as long as they add up to zero.

When the spins are coupled to an external magnetic field and the
temperature is sufficiently small, the freezing of the chiralities has
a strong effect on the low-temperature spin properties.  First of all,
the equilibrium magnetization for small magnetic fields has a clear
non-linear behavior, whereas for large enough $h$, we obtain $m \propto h$. The
actual magnetic field range for which the non-linear behavior is
detected strongly depends upon the temperature.  
However, the non-linearity cannot be meaningfully casted into 
a divergent $\chi_3$, in contrast to traditional spin glasses. 
Still, very strong
non-linear effects in the magnetization curves are clearly present
and point to a spin-glass reminiscent behavior. Secondly, there is a
clear difference between the FC and the ZFC measurements of the
magnetization, typical of spin-glass models. A difference with
ordinary spin-glass systems, representing an original aspect of our
results is that the characteristic temperature $T_m$ below which the
slow spin response takes place does not coincide with $T^*$ and strongly
depends on the external magnetic field, i.e., $T_m=T_m(h)$.  On the
other hand, the freezing transition of an ordinary spin glass
corresponds to a thermodynamic transition and the freezing temperature
is independent of the particular protocol.  Nonetheless, the very slow
spin response to an external magnetic field is typical of ordinary
out-of-equilibrium systems, and here it is driven by the freezing of
the chiralities.  In this respect, the richness of our model consists
in having two distinct temperatures, i.e., $T^*$, which characterizes
the entropic selection of chirality configurations, and $T_m(h)$, which
characterizes the establishment of the slow spin dynamics. On the
contrary, in the ordinary spin-glass models, there is only one
characteristic temperature.

So the results we have obtained on the classical
version of the effective model of the S=1/2 Heisenberg model on the
trimerized {\it kagome} clearly show that the interplay of two 
local degrees of freedom can lead to a very slow dynamics, and to
a behavior which is in many respects reminiscent of the spin-glass
phenomenology. One point worth noticing at that stage is that this
behavior is, as far as we can tell, independent of the dynamics chosen
in the numerical simulations. In particular, we have performed a few
simulations with simultaneous updates of spins and chiralities at the
same site, and the results were unchanged. This is a clear improvement
over the simplified model of Ref.~\onlinecite{miladean}, in which the 
entropic barriers were related to the dynamics. The present results
suggest that the development of a very slow dynamics is actually
a generic feature of models with two local degrees of freedom.

Coming back to the experimental evidence in favor of a spin-glass 
behavior in ${\rm SrCr_{8-x}Ga_{4+x}O_{19}}$, the present
results are very promising in several respects. First of all,
the model was motivated by physical considerations of the 
exchange processes in this system, in which the {\it kagome} layers
are trimerized. Besides, the broken ergodicity which is at the origin
of the very slow dynamics is neither related to a specific dynamics
nor to disorder but to the presence of two local degrees of freedom,
a very natural consequence of the special topology of the trimerized
{\it kagome} lattice. In particular, although we started from spins 1/2,
all half-integer spin models on the trimerized {\it kagome} lattice
lead to very similar effective models with
two local degrees of freedom.~\cite{note}
In addition, the characteristic temperature at which large time
scales appear is much lower than the coupling constant, again
in agreement with ${\rm SrCr_{8-x}Ga_{4+x}O_{19}}$. 
Finally, the spins have a very anomalous response to magnetic field,
with a clear difference between field-cooled and zero-field-cooled protocols,
while retaining a very fast dynamics at low temperatures. As far as
we know, this is a unique feature of the present model, which could 
be related to the muon spin rotation observation that
only a tiny part of the spins are actually frozen in 
${\rm SrCr_{8-x}Ga_{4+x}O_{19}}$ below the spin-glass 
transition.~\cite{uemura}

There are clear differences however between our model and the properties
of ${\rm SrCr_{8-x}Ga_{4+x}O_{19}}$. An obvious one is the specific
heat, which is quadratic below the spin-glass transition,
whereas in our model it goes to a constant. This behavior is
clearly a consequence of our approximation to treat spin and
chirality degrees of freedom as classical variables. Given the
quasi-two dimensionality of the compound, one can speculate
that a quantum treatment would lead to a $T^2$ behavior of
the specific heat for the quadratic degrees of freedom, in agreement
with experiments. This could actually be a natural explanation
of this behavior, which is {\it not} the most common behavior
in spin glasses, which tend to have a linear specific heat at
low temperature.
At that stage, the main discrepancy between the phenomenology 
described in the present paper and the experimental results obtained
in ${\rm SrCr_{8-x}Ga_{4+x}O_{19}}$ is the observation of a clear phase
transition in this compound with a divergent non-linear susceptibility.
It is likely that a true phase transition would require to include
inter-layer coupling, like in disordered models of spin-glasses.
Whether including such a coupling can produce a divergent non-linear
susceptibility is left for future investigation. 
We should also note
that the effective model for spins 1/2 has specific features with 
respect to the effective models for larger half-integer spins.
In particular, the identification of exact ground states does not
seem to be possible for larger spins,~\cite{note} and the 
phenomenology of these models might turn out to be different in
some respects, although we are confident that the presence of two
degrees of freedom will still lead to an anomalous response of the
spins to an external magnetic field and to the development of
large time scales.

\acknowledgments

It is a pleasure to thank J.-P. Bouchaud, D. Dean, P. Delos Rios, S. Franz,
M. Mambrini, M. Sellitto, 
F. Ricci Tersenghi, and M. Zhitomirsky for enlightening discussions.  This work
has been supported by the Swiss National Fund and by Istituto
Nazionale per la Fisica della Materia (INFM).

\appendix
\section{Monte Carlo technique}\label{montecarlo}

In this Appendix, we describe the Monte Carlo method used throughout
our study. For simplicity, let us consider the Hamiltonian:
\begin{equation}
  {\cal H} = \sum_{\langle i,j \rangle}
  J_{ij} \, {\vec \sigma}_i \cdot {\vec \sigma}_j,
\end{equation}
where ${\vec \sigma}_i$ are classical unit spin vectors on the sites
of an $N$-site lattice and $J_{ij}$ are magnetic couplings.

In order to study the dynamics of this Hamiltonian, we have performed
Monte Carlo simulations using the heat-bath algorithm to update the
vector directions.~\cite{young} Each spin is assumed to be in contact
with a heat bath and is immediately put into a local equilibrium with
respect to the instantaneous effective field on it from the nearest
neighbor spins:
\begin{equation}
  {\vec H}_i = \frac{1}{T} \sum_{{\langle j \rangle}_i}
  J_{ij} \, {\vec \sigma}_j,
\end{equation}
where ${\langle j \rangle}_i$ indicates the nearest neighbors of the
site $i$ and $T$ is the temperature.

The new direction of a given spin is determined by a polar angle
$\theta$ and an azimuthal angle $\phi$ relative to ${\vec H}_i$,
through the following Boltzmann distribution:
\begin{equation} \label{eq:dist}
  P(\cos{\theta}) = \frac{H}{2 \sinh{H}} \, e^{H \cos{\theta}},
\end{equation}
where $H = |{\vec H}_i|$. The two angles $\theta$ and $\phi$ are
chosen with the use of two random numbers $R_1$ and $R_2$, uniformly
distributed between $0$ and $1$.  Even though ~(\ref{eq:dist}) is
independent of $\phi$, it is useful to update this variable in order
to increase the rate at which the phase space is sampled. We therefore
chose a new $\phi$ with
\begin{equation}
  \phi = 2 \pi R_2.
\end{equation}
The polar angle $\theta$ is instead obtained by
\begin{equation}
  \cos{\theta} = \frac{1}{H} \ln [1 + R_1 (e^{2 H} - 1)] - 1.
\end{equation}
It is important to note that the heat bath method does not rely on an
accept-reject approach. Instead, the new configuration is already
generated with the right probability.

When both spins and chiralities are present in the Hamiltonian, the
same updating algorithm can be applied to the chiralities.  In our
simulations, we consecutively chose a random site on the lattice,
updated the spin and then the chirality on another random site.
Quantities like the energy or the magnetization where usually
calculated after $N$ such moves, i.e., after the system had undergone
one Monte Carlo step per site on average.

Because we are working with continuous degrees of freedom and not with
discrete variables as in the Ising model for example, the heat bath
algorithm is better suited than the conventional Metropolis method. In
the Metropolis method, at low temperatures, one starts rejecting many
moves because the system prefers to make small fluctuations about
local minima of the energy.  Thus, it may be very difficult to reach
equilibrium or even to know whether one has reached it. By contrast,
the heat bath method provides an efficient way to reach the actual
minimum of the energy because a new vector orientation is obtained
every time, at any temperature.



\begin{thebibliography}{99}
\bibitem{note} An effective Hamiltonian, similar to the one of
  Ref.~\onlinecite{mila} has been derived for all generic half-integer 
  spins $S$, and
  the analysis in this direction is in progress (M. Mambrini, private
  communication).
\end{thebibliography}
\end{document}